\documentclass[aip,jmp,amsmath,amssymb,preprint,reprint]{revtex4-1}

\usepackage{graphicx}
\usepackage{dcolumn}
\usepackage{bm}

\begin{document}

\preprint{AIP/123-QED}

\title[Sample title]{The  principle of superposition for waves}

\author{L. M. Ar\'{e}valo Aguilar}
 \email{olareva@yahoo.com.mx}
\author{C. Robledo-S\'anchez}
\author{M. L. Arroyo Carrasco}
\author{M. M. Mendez Otero}
\affiliation{ Facultad de Ciencias F\'isico Matem\'aticas\\
Benem\'erita Universidad Aut\'onoma de Puebla, 18  Sur y Avenida San Claudio, Col. San Manuel.\\
C.P. 72520, Puebla, Pue. M\'{e}xico}


\date{\today}

\begin{abstract}
In this paper we will argue that the superposition of waves can be calculated and taught in a simple way. We show, using the Gauss's method to sum an arithmetic sequence, how we can 
construct the superposition of waves - with different frequencies - in a simple conceptual way that it is easy to teach.
By this method we arrive to the usual result where we can express the superposition of waves as the product of factors, one of them
with a cosine funtion where the argument is the average frequency. Most important, we will show that the superposition of waves with sligthly different frequencies produces a \emph{phase modulation} thogether with an amplitude moldulation. It is important to emphasize this result because, to the best of our knowledge, almost all textbooks only mention that there is an amplitude modulation.
\end{abstract}

\pacs{Valid PACS appear here}


\maketitle

\section{\label{sec:level1}Introduction}

Superposition of waves is one of the fundamental concepts in the subject matters of Waves \cite{crawford,pain,main,french}
and Optics \cite{hecht, guenther}. It allows the study and explanation of interference patterns in Optics, which is
one of the most beautiful physical phenomenon. Futhermore, it allows physicists
 to explain many interesting sound phenomena like beat phenomenon, which is a slow variation in the intensity
of sound when we add two waves of sligthly different frequencies. The beat phenomenon is used by musicians in
tunin theirs musical instruments \cite{breazeale}, it is caused by the amplitude modulation effect produced by superposing waves of sligthly different frequencies. The conceptual foundation in which is based the study 
of these phenomena is the Principle of Superposition, which rougly speaking state, in the case of waves, that the resultant of adding $N$
waves is the sum of the individual waves. 

When this subject is taught in the beginning years of undergraduate physics or engineering courses\cite{halliday,fishbane},  the analysis usually is carried out by summing only two waves as follows: 
\begin{equation}
y(x,t)=A_1 \cos\left(\omega_1t-k_1x+\phi_1\right)+A_2\cos\left(\omega_2 t-k_2x+\phi_2\right).\label{ecu1}
\end{equation}

If the waves are in phase (i. e. $\phi_1=\phi_2$) and with the same amplitude, i.e. $A_1=A_2=A$ (and observed at some fixed point), we can express the result as the multiplication of two sinusoidal waves as follows \cite{crawford}:
\begin{equation}
y(t)=2A \cos\left(\frac{(\omega_2+\omega_1)t}{2}\right)\cos\left(\frac{(\omega_2-\omega_1)t}{2}\right). \label{two-waves}
\end{equation}
If the two frequencies are rather similar, that is when:
\begin{equation}
\omega_2 \approx \omega_1, \label{approx1}
\end{equation}
then, it is stated in many texbooks that equation (\ref{two-waves}) represents a wave that oscillates at frequency $(\omega_2+\omega_1)/2$ and whose intensity increase and decrease at the beating frequency $(\omega_2-\omega_1)$. In fact, it is common to write Equation (\ref{two-waves}) in the following way:
\begin{equation}
y(t)=A(t)\cos\left(\frac{(\omega_2+\omega_1)t}{2}\right), \label{two-waves-2}
\end{equation}
where $A(t)=2A \cos\left((\omega_2-\omega_1)t/2\right)$. Equation (\ref{two-waves-2}) is interpreted as a wave that oscillates at the high frequency $(\omega_2+\omega_1)/2$ multiplied by a slowly variable amplitude $A(t)$. Therefore, most texbooks agree that Equation (\ref{two-waves-2}) clearly shows that the sum of two waves of almost equal frequency produces the phenomenon of amplitude modulation.

Accordingly, texbooks reproduce the plot of Equation (\ref{two-waves}) when $\omega_2 \approx \omega_1$. However, in almost all textbooks, nothing is mentioned about the case when equation (\ref{approx1}) is not fulfilled. Futhermore, maybe because to sum $N$ waves  usually is carried out using complex functions, the next step which consist in the task of summing $N$ waves is not done in the beginning courses.

In most universities, this next step in the study of superposition of waves (i. e. suming $N$ waves for $N>2$) arrives with the course called Oscillation and Waves. Then, representing waves as complex functions (i.e. $e^{i\omega t}$) it is possible to obtain the following analitycal expression for the sum of $N$ waves of equal succesive phase diference $\delta\omega$ \cite{pain,crawford}:  
\begin{equation}
	y(x,t)=A\cos\left(\frac{\omega_N+\omega_1}{2}t\right)\frac{\sin \left(N\delta\omega t/2\right)}{\sin(\delta\omega t)}, \label{suma-seno}
\end{equation}
where $\omega_N$ is the last frequency of the sequence and $\omega_1$ is the first. In general, the $N-th$ frequency can be obtaining by summig $N-1$ times $\delta\omega$ to the wave with frequency $\omega_1$, that is:
\begin{equation}
\omega_N=\omega_1+(N-1)\delta\omega, \label{omega-N}
\end{equation}

To deduce Equation (\ref{suma-seno}), the textbook's authors sum the geometric progression $S(z)=1+z+z^2+\cdots+z^N$, where $z=e^{i\delta\omega}$, see pages $29$ and $30$ in reference \cite{pain} or page $287$ in reference \cite{crawford}. It is worth to mention that another way to arrive to Equation (\ref{suma-seno}) is by using the phasor method \cite{crawford,pain}. Also, it is important to highly that in the same way that occur with Equation (\ref{two-waves}) most textbooks usually plot Equation (\ref{suma-seno}) when 
\begin{equation}
\delta\omega\ll\omega_1,\label{delta-approx}
\end{equation}
which implies (for $N$ not so high):
\begin{equation} 
\omega_N\approx\omega_1. \label{approx-N} 
\end{equation}
Equation (\ref{approx-N}) is pretty similar approximation than Equation (\ref{approx1}). Both plots, that of Equations (\ref{two-waves}) and (\ref{suma-seno}), are practically the same when conditions given by Equations (\ref{approx1}) and (\ref{approx-N}) are fulfilled. The oscillations in both situations are similar, and it is taken for granted that the plot of Equation (\ref{suma-seno}) oscillates at frequency $(\omega_N+\omega_1)/2$ with the envelope oscillating at frequency $(\omega_N-\omega_1)/2$.

To summarize, almost all texbooks on the subject matter of Waves \cite{crawford,pain,main,french} and Optics \cite{hecht, guenther} agree that when two (or $N$) waves of sligthly different frequencies are superposed the most prominent observed effect is the \textit{amplitude modulation}. Almost all of them does not mention any thing about \textit{phase variation or modulation}.

Because of being such an important and fundamental concept, the study of the superposition of waves is of high importance and deserves its study since the beginning of the syllabus to make the most of learning. Therefore, it would be convenient to find a way to introduce the study of these concepts in full details since the beginning. In particular, as probably many students does not know the subject of complex numbers, it would be convenient to present at this stage the superposition of $N$ waves, when $N>2$, without the use of complex functions. 

On the other hand, from the point of wiew of Physical Education Research (PER) there is the goal to present new conceptual ways to approach the teaching of relevant physical concepts. Then, reaching this goal is one of the most important learning steps that a teacher can focuses on. It is worth to emphazise that one of the fundamental goals of the PER's field is to discover new and imaginative ways of teaching physics. Part of this goal could be accomplished by finding easier conceptual and mathematical ways to solve model problems.

In this work, we present a method to sum $N$ waves in a simple conceptual way; just by using the trick, usually attributed to Gauss, of arranging numbers by pairs to sum them. This is a quite simple conceptual  method that could be taught to any student with only basic knowledge on trigonometric identities and, of course, it could be taught to more advanced students. Also, we analyze this superposition in other settings different to the usual restriction of $\delta\omega\ll\omega_1$. We have found that the sum of $N$ waves is more richer and complex phenomenon than that presented in some texbooks. The most important discovery of this work is that\textit{ we found that together with an amplitude modulation there exist a phase modulation when we superpose two or more waves of slightly different frequencies}. It is worth to highly this result because it could have important practical implications.

Furthermore, this approach to study the superposition of waves allows to introduce an historical account in teaching, both to increase the student interest and to show the problems associated with the history of scientific facts, in this case the problem associated with who (and how) discovered the procedure to sum an arithmetic sequence. Also, there is an interesting question in the work done by Raleigh that we will draw at the end of subsection 3 A. There have been many advocates about the convenience of treating problems of historical science facts in the classroom to enhance the grasp of something that is complicated or difficult to understand, see for instance the  the recent published review by Teixeira, et. al. \cite{teixeira}, the illuminating works of M. A. B. Whitaker \cite{whitaker,whitaker2} and see, also, reference \cite{abou}.

\section{Arithmetic Sequence}

Whether apocryphal or not it is well known the story of Gauss's method to sum an arithmetic sequence of numbers \cite{hayes} ( or Alcuin of York method \cite{hayes} or Archimedes \cite{hayes2}). The story and its weakness and drawbacks are well summarized in the article by Brian Hayes \cite{hayes,hayes2}. When Gauss was a child his teacher gave all the students the task of summing an arithmetic progression \cite{hayes}; in order to be explicit we suppose here that the arithmetic sequence was the first one hundred numbers. It is said that using his great talent Gauss didn't make the procedure of adding one number to the other, instead he grouped together the numbers in pairs and found that the sum of a pair of numbers was the same for numbers located at equidistan distance from the extrem, that is $100+1$=$99+2$=$98+3$=$\cdots$=$52+49$=$51+50=101$. Then, the sum is $50$ times $101$. It is worth to mention that 
the difference between the grouped numbers is lowered by two, given a sucession of odd numbers,
that is $100-1=99,$ $99-2=97$, $98-3=95$, $\cdots$, $52-49=3$, $51-50=1$.

Also, you can choose anhother sequence and apply the same rule. For example, the sequence $0.1,0.2$,$\cdots$ $0.9,1$, and group it in the same way $1+0.1=0.9+0.2=\cdots=0.6+0.5=1.1$. Furthermore, $1-0.1=0.9, 0.9-0.2=0.7, \cdots, 0.6-0.5=0.1.$ Even you can use the same procedure to sum distances; this is not a surprise if we consider that to every real number $r$ correspond a distance given by the
absolute value $\left|r\right|$.

More generally, an arithmetic sequence is a set of numbers such that each of them is obtained by summing to the previous
number a constant, say $\delta\omega$. For example, the set $(1, 11, 21, 31, \cdots, 101)$ is an arithmetic progresion with constant $\delta\omega=10$, which has as a first number $1$ and as a last number $101$. 

An arbitrary term of the arithmetic sequence can be calculated as follows: If $a_1$ is
the first term then the secon term $a_2=a_1+\delta\omega$ is:
\begin{equation}
a_1+(2-1)\delta\omega
\end{equation}
the third term can be written as:
\begin{equation}
a_1+(3-1)\delta\omega,
\end{equation}
therefore, the $i-th$ term of the arithmetic sequence is given by:
\begin{equation}
a_i=a_1+(i-1)\delta\omega. \label{N-th}
\end{equation}

As was stated above, an arithmetic sequence has the property that the sum of two numbers, which are at the same distance (equidistant) repectively from its nearest extrem in the sequence, is equal. To prove this, suppose that $a$ is the first number of the sequence and $l$ the last number, then if $x$ is the number located at the $i-th$ point from the first and $y$
is the number located at the $i-th$ point from the last then:
\begin{eqnarray}
x=a+(i-1)\delta\omega\\
y=l-(i-1)\delta\omega
\end{eqnarray}
therefore
\begin{equation}
x+y=a+l. \label{arithmetic-equality}
\end{equation}

Equation (\ref{arithmetic-equality}) shows that the sum of two elements in the arithmetic sequence that are located at equidistant points from the extrems (i. e. one of them from the first number in the sequence and the other from the last) is the same.

On the other hand, it is worth to notice that Equation (\ref{N-th}) is the same Equation (\ref{omega-N}) (and is the same that Equation (\ref{condition-arithmetic}) below), this will allow us to think about the sum of $N$ waves as representing a superposition of an arithmetic sequence of frequencies and to use the properties of this kind of sequences, in particular the property given by Equation (\ref{arithmetic-equality}), to sum them. This way of thinking contrast with that of textbook's authors who used to think about the sum of $N$ waves in terms of geometric sequences.

The historical problem about who was the first person who discovered the trick to sum an arithmetic sequence is explained by Haynes in references \cite{hayes} and \cite{hayes2}. This historical sucess can be used to show students the problems to exactly determine some historical facts and to increase their understanding of the nature of science \cite{teixeira}. It is notable that many people really know that the trick was found by Gaus, but few people know that such trick was published before (in books) by at least two people: Alcuin of York and Archimedes \cite{hayes,hayes2}.  

\hspace{2cm}

\section{\label{sec:level2}Superposition of waves with different frequencies at a fixed point}

\subsection{The sum of frequencies as a constant}
As was stated above, when we add two waves which have slightly different frequencies their superposition produces an amplitude-modulated wave. Based on the superposition principle, the sum of two or more waves is a wave too, which satisfaces the wave equation. The magnitude of the wave sum at any space point and at some time depends on the phase value of every wave component. That is, the wave's phase is a function of both the space and time.

In this section we are going to study the case where we superpose $N$ waves (with $N$ an even number) with the same phase constant, same amplitude and different frequencies at a fixed point $x$. We consider an interval of frequencies $\Delta\omega$ and make a partition
of the interval in an even number between, say,  $\omega_1$ to $\omega_{100}$, i. e. $\Delta\omega=\omega_{100}-\omega_1$. That is, we have:
\begin{equation}
	y_T(x,t)=A\sum_{i=1}^{N=100}\cos(k_ix-\omega_it+\phi_i),\label{suma}
\end{equation}
where $\omega_i$ is the angular frequency of the $i-th$ wave, $k_i$ is the wave number, and $\phi_i$ is the initial phase constant. For the sake of simplicity we are going to consider the case where the waves have the same initial phase constant. Also, as we are interested in analyzing the case where the waves are superposed at some fixed point, without loss of generality we can set $x=0$. Therefore Equation (\ref{suma}) reduces to:

\begin{equation}
	y_T(t)=A\sum_{i=1}^{N=100}\cos(\omega_it).\label{sumax0}
\end{equation}
We consider the case when the upper limit in the sum, i.e. $N$, is an even number. Also, we make the partition of the $\Delta\omega$ by adding a $\delta\omega$ to the next frequency in the sequence, that is, we have:
\begin{equation}
\omega_N=\omega_1+(N-1)\delta\omega. \label{condition-arithmetic}
\end{equation}

 Therefore, if we compare Equation (\ref{condition-arithmetic}) with Equation (\ref{N-th}) we conclude that the set of frequencies form an arithmetic sequence. Therefore, in order to be able to use Gauss method, we group the sum given in Equation (\ref{sumax0}) by pairs as follows:
 
\vspace{.5cm}
\begin{widetext}
\begin{eqnarray}
	y_T(t)=A\bigg\{\left[\cos(\omega_{100}t)+\cos(\omega_1t)\right]+\left[\cos(\omega_{99}t)+\cos(\omega_2t)\right]+\cdots+
	\left[\cos(\omega_{51}t)+\cos(\omega_{50}t)\right]\bigg\}.\label{groupingsum}
\end{eqnarray}
\end{widetext}
Now, using the following trigonometric identity
\begin{eqnarray}	&&\cos(\omega_{i}t)+\cos(\omega_jt)=\nonumber\\&&2\cos\left(\frac{(\omega_i+\omega_j)t}{2}\right)\cos\left(\frac{(\omega_i-\omega_j)t}{2}\right),\label{trigoidentity}
\end{eqnarray}
we can rewrite the sum of the grouped cosine functios in Equation (\ref{groupingsum}) as the product of two cosines functions. Then, by
substituting Equation (\ref{trigoidentity}) in Equation (\ref{groupingsum}) we have:
\vspace{.5cm}
\begin{eqnarray}
y_T(t)&=&A\Bigg\{\left[2\cos\left(\frac{(\omega_{100}+\omega_1)t}{2}\right)\cos\left(\frac{(\omega_{100}-\omega_1)t}{2}\right)\right]+ \nonumber\\	&&\left[2\cos\left(\frac{(\omega_{99}+\omega_2)t}{2}\right)\cos\left(\frac{(\omega_{99}-\omega_2)t}{2}\right)\right]+\cdots\nonumber\\ 	&+&\left[2\cos\left(\frac{(\omega_{51}+\omega_{50})t}{2}\right)\cos\left(\frac{(\omega_{51}-\omega_{50})t}{2}\right)\right]\Bigg\}.\label{substrigo}
\end{eqnarray}
\vspace{.5cm}
Now, as by Equation (\ref{condition-arithmetic}) $\omega_{100}=\omega_1+99\delta\omega$,  $\omega_{99}=\omega_1+98\delta\omega$, and so
 on, therefore $\omega_{100}+\omega_{1}=\omega_{99}+\omega_{2}=\cdots=\omega_{51}+\omega_{50}=2\omega_1+99\delta\omega$, then
\vspace{.5cm}
\begin{eqnarray}
  \cos\left(\frac{(\omega_{100}+\omega_1)t}{2}\right)=\cos\left(\frac{(\omega_{99}+\omega_2)t}{2}\right)=\cdots\nonumber\\
  =\cos\left(\frac{(\omega_{51}+\omega_1)t}{2}\right)=\cos\left(\frac{(2\omega_1+99\delta\omega)t}{2}\right),
\end{eqnarray}
so, we can factorize the cosine of the frequencies' sum in equation (\ref{substrigo}) as follows:
\begin{eqnarray}
	&&y_T=2A
	\cos\left(\frac{(2\omega_1+99\delta\omega) t}{2}\right)\Bigg\{\cos\left(\frac{(\omega_{100}-\omega_1)t}{2}\right)+\nonumber\\
	&&\cos\left(\frac{(\omega_{99}-\omega_2)t}{2}\right)+\cdots+
	\cos\left(\frac{(\omega_{51}-\omega_{50})t}{2}\right)\Bigg\},\label{sum-factorizado}
\end{eqnarray}
on the other hand, by Equation (\ref{condition-arithmetic}), we have $\omega_{100}-\omega_{1}=99\delta\omega,\omega_{99}-\omega_{2}=97\delta\omega,\cdots,\omega_{51}-\omega_{50}=\delta\omega$. Then, the Equation (\ref{sum-factorizado}) can be written as:
\begin{eqnarray}
	y_T=2A
	\cos\left(\frac{(2\omega_{1}+99\delta\omega)t}{2}\right)\Bigg\{\cos\left(\frac{99\delta\omega t}{2}\right)+\nonumber\\
	\cos\left(\frac{97\delta\omega t}{2}\right)+\cdots+
	\cos\left(\frac{\delta\omega t}{2}\right)\Bigg\}.\label{sum-factoriz-delta}
\end{eqnarray}

We can still carry out a further step and group together by pairs the cosine functions that are inside the braces of Equation (\ref{sum-factoriz-delta}). By applying Equation (\ref{trigoidentity}), this will produce twenty five factors with the common factor $\cos(25\delta\omega t)$ which can be factorized. You can make additional steps by grouping by pairs the rest of the cosine fucntions, and you can follow this procedure until you arrive to a short expression. We left this as an exercise to the students. To show how the procedure works in full details, in the next subsection we carry out all the steps in the case where we superpose ten waves.
  
\begin{figure*}
\includegraphics{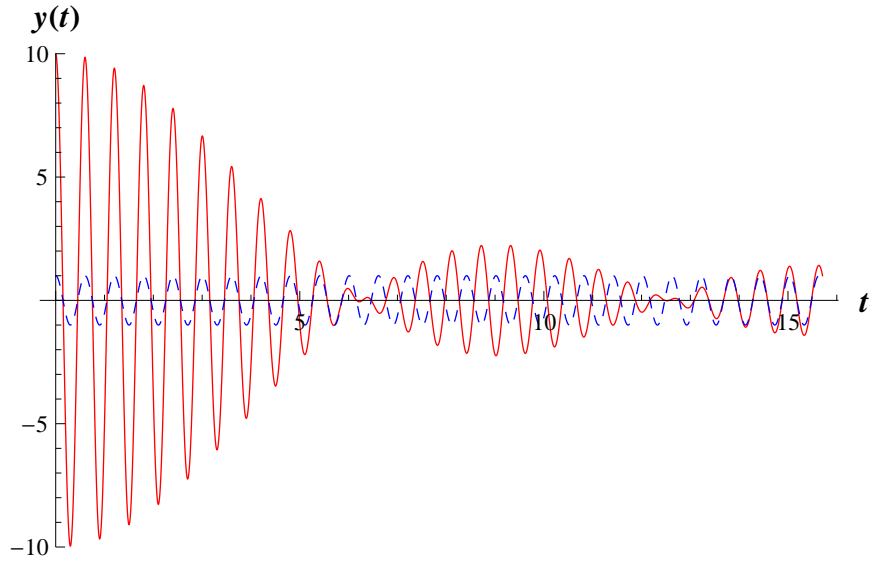}
\caption{Plot of equation (\ref{final-sum}) for $\delta\omega\ll \omega_1$. We have set $\omega_1=10$, A=1, and $\delta\omega=\omega_1\times10^{-2}$. The dotted line is the plot of the function $\cos(\bar{\omega}t).$}
\end{figure*}

\subsubsection{Example 1}
In this subsubsection we are going to study the case of the superposition of ten waves. In this case the arithmetic sequence is $\omega_1, \omega_{2}=\omega_1+\delta\omega, \omega_{3}=\omega_1+2\delta\omega, \omega_{4}=\omega_1+3\delta\omega, \omega_{5}=\omega_1+4\delta\omega, 
\omega_{6}=\omega_1+5\delta\omega, \omega_{7}=\omega_1+6\delta\omega, \omega_{8}=\omega_1+7\delta\omega, \omega_{9}=\omega_1+8\delta\omega,
 \omega_{10}=\omega_1+9\delta\omega$. The total superposition will be, where we have grouped together the cosines funcions by pairs in a convenient way:
 
\begin{eqnarray}	y_T(t)=A\bigg\{&&\left[\cos(\omega_{10}t)+\cos(\omega_1t)\right]+\left[\cos(\omega_{9}t)+\cos(\omega_{2}t)\right]+\nonumber\\
&&\left[\cos(\omega_{8}t)+\cos(\omega_{3}t)\right]+\left[\cos(\omega_{7}t)+\cos(\omega_{4}t)\right]\nonumber\\
&&+\left[\cos(\omega_{6}t)+\cos(\omega_{5}t)\right]\bigg\}.\label{}
\end{eqnarray}
 
 Therefore the total sum, as $\omega_{10}+\omega_{1}=\omega_{9}+\omega_{2}=\omega_{8}+\omega_{3}=\omega_{7}+\omega_{4}=\omega_{6}+\omega_{5}=2\omega_1+9\delta\omega$ and $\omega_{10}-\omega_{1}=9\delta\omega, \omega_{9}-\omega_{2}=7\delta\omega, \omega_{8}-\omega_{3}=5\delta\omega, \omega_{7}-\omega_{4}=3\delta\omega, \omega_{6}-\omega_{5}=\delta\omega$, is:
\begin{eqnarray}
y_T(t)=2A	&&\cos\left[\frac{(2\omega_1+9\delta\omega)t}{2}\right]\Bigg\{\cos\left[\frac{9\delta\omega t}{2}\right]+
	\cos\left[\frac{7\delta\omega t}{2}\right]+\nonumber\\
	&&\cos\left[\frac{5\delta\omega t}{2}\right]
	+\cos\left[\frac{3\delta\omega t}{2}\right]+\cos\left[\frac{\delta\omega t}{2}\right]\Bigg\}.\label{10-sum-factory}
\end{eqnarray}

The next step is to group together by pairs the cosine functions that are inside the braces of Equation (\ref{10-sum-factory}) and to use the trigonometric identity given by Equation (\ref{trigoidentity}) as follows:
\vspace{.5cm}

\begin{widetext}
\begin{eqnarray}
\Bigg\{\left[\cos\left(\frac{9\delta\omega t}{2}\right)+\cos\left(\frac{\delta\omega t}{2}\right)\right]+ \left[\cos\left(\frac{7\delta\omega t}{2}\right)+\cos\left(\frac{3\delta\omega t}{2}\right)\right]+
\cos\left(\frac{5\delta\omega t}{2}\right)	\Bigg\}=\nonumber\\
\Bigg\{2\cos\left(\frac{5\delta\omega t}{2}\right)\cos\left(\frac{4\delta\omega t}{2} \right)+2\cos\left(\frac{5\delta\omega t}{2}\right) \cos\left(\frac{2\delta\omega t}{2}\right)+ \cos\left(\frac{5\delta\omega t}{2}\right)\Bigg\}=\nonumber\\
\cos\left(\frac{5\delta\omega t}{2}\right)\Bigg\{2\left[\cos\left(2\delta\omega t\right)+\cos\left(\delta\omega t\right)\right]+1\Bigg\}=
\cos\left(\frac{5\delta\omega t}{2}\right)
\Bigg\{4\cos\left(\frac{3\delta\omega t}{2} \right)\cos\left(\frac{\delta\omega t}{2}\right)+1\Bigg\}.\label{total-factory1}
\end{eqnarray}
\end{widetext}

\begin{figure*}
\includegraphics{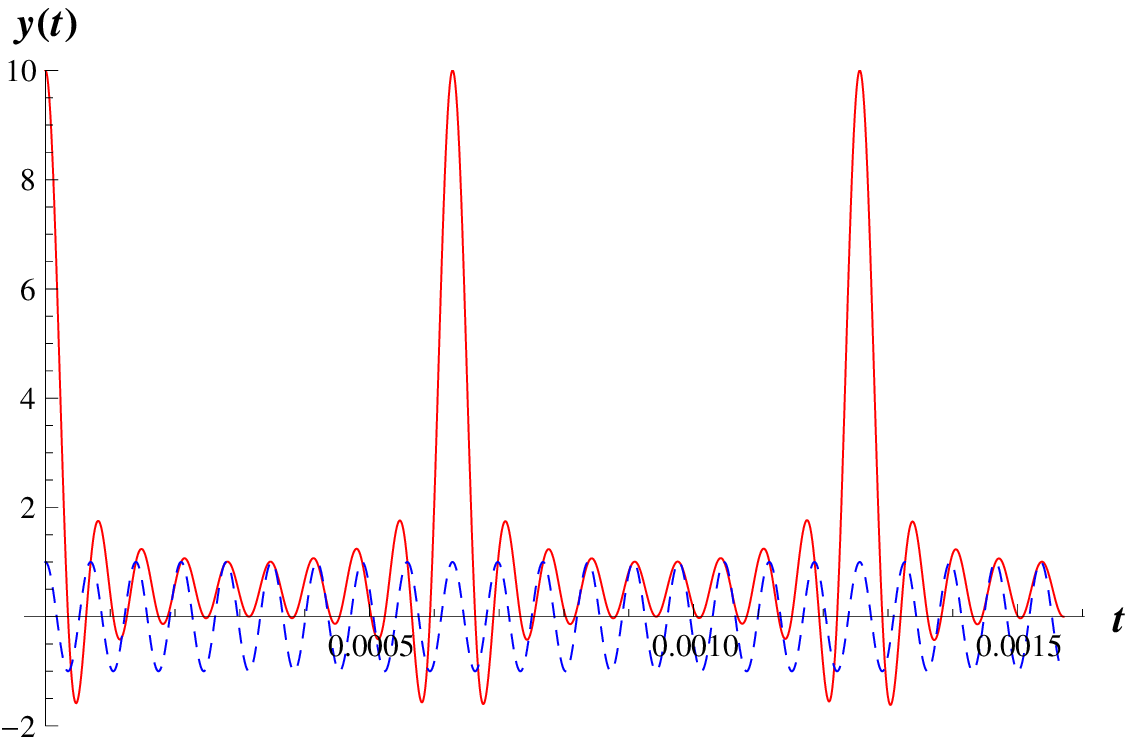}
\caption{Plot of equation (\ref{final-sum}) for $\delta\omega\gg\omega_1$. We have set $\omega_1=10$, $A=1$, and $\delta\omega=\omega_1\times10^3$. The dotted line is the plot of the function $\cos(2\bar{\omega}t)$.}
\end{figure*} 
substituting Equation (\ref{total-factory1}) into Equation (\ref{10-sum-factory}) we arrive to:
\begin{widetext}
\begin{eqnarray}
y_T(t)=2A
	\cos\left[\frac{(2\omega_1+9\delta\omega)t}{2}\right]\cos\left(\frac{5\delta\omega t}{2} \right)
	\times\Bigg\{4\cos\left(\frac{3\delta\omega t}{2} \right)\cos\left(\frac{\delta\omega t}{2}\right)+1\Bigg\}.\label{final-sum}
\end{eqnarray}
\end{widetext}
Equation (\ref{final-sum}) has, like Equation (\ref{suma-seno}), a cosine term with an argument that oscillates at the mean frequency $\overline{\omega}=(\omega_{10}+\omega_{1})/2=(2\omega_1+9\delta\omega)/2$. The plot of Equation (\ref{final-sum}) when $\delta\omega\ll\omega_1$ is the same as the plot of Equation (\ref{suma-seno}) under similar condition. In this case, that is when Equation (\ref{delta-approx}) is fulfilled, the function seem effectively to oscillate at the mean frequency $\overline{\omega}=(2\omega_1+9\delta\omega)/2$. However, when we plot Equation (\ref{final-sum}) togheter with the cosine function that oscilates at the mean frequency $\bar{\omega}$ we find that there is a phase delay each time that there is a destructive interference. Whe show this in Figure 1, where we plot Equation (\ref{final-sum}) and also we have plotted the cosine function that oscillates at the mean frequency, i. e.  $\cos{(\overline{\omega}t)}$. The plot clearly shows that there is a phase delay between the superposition of waves given by Equation (\ref{final-sum}) and the $\cos{(\overline{\omega}t)}$ function. To the best of our knowledge, this phase delay was not previously noticed or reported. Also, \textit{it is worth to mention that the same phase delay is observed if we add two, ten or more waves}. We will give a probable explanation of this effect in the last part of this subsubsection. Previously, we are going to examine Equation (\ref{final-sum}) in a different setting.

On the other hand, we can explore the effects produced by different conditions on the superposition of waves. In particular, we can focuses in the condition given by:
\begin{equation}
\delta\omega\gg\omega_1. \label{conditiongg}
\end{equation}

As $\omega_N+\omega_1=2\omega_1+(N-1)\delta\omega$ and $\omega_N-\omega_1=(N-1)\delta\omega$, then condition given by Equation (\ref{conditiongg}) implies:

\begin{equation}
\omega_N+\omega_1\approx\omega_N-\omega_1.
\label{}
\end{equation}

In Figure 2 we plot Equation (\ref{final-sum}) when condition given by  Equation (\ref{conditiongg}) 
is fulfilled.  In this case it is clear that the wave sum does not oscillates at the mean frequency $\overline{\omega}$, this is show in Figure 2 where we plot also the function $\cos(2\bar{\omega})$. It is easy to see that the wave sum oscillates at a frequency similar to $2\bar{\omega}=\omega_{10}-\omega_1$; however, there are deviations from $\cos(2\bar{\omega})$ which strongly suggest that this plot shows a frequency or phase modulation. This could explain the phase delay show in Figure 1 which then could be produced by a phase modulation effect.

To state it clearly: \textsl{together with an amplitude modulation, in the superposition of waves with similar frequencies, it is probable that there is also a phase modulation. Which in the case of Figure 1 it is expressed by a phase delay}.

To show that this is effectively the case, let us recall that a phase modulated signal $s_m(t)$ is represented by the following equation, see problem 6.30 in Crawford's book \cite{crawford}:
\begin{equation}
s_{pm}(t)=A_{pm}\cos{\left[\omega_ct+ a_m\sin{(\omega_m t)}\right]}. \label{ph-equation}
\end{equation}

Using basic trigonometric identities, Equation (\ref{ph-equation}) can be rewritten in the following form:
\begin{equation}
s_{pm}(t)=A_{pm} \cos{(\omega_c t)}\cos{\left(\theta(t)\right)}-A_{pm}\sin{(\omega_c t)}\sin{\left(\theta(t)\right)},\label{ph-equ-expan}
\end{equation}
where $\theta(t)=a_m\sin{(\omega_m t)}$. Now, if we consider the case where 
\begin{equation}
\left|\theta(t)\right|\ll1,\label{theta-cond}
\end{equation} 
then $\cos{\theta(t)\approx 1}$ and $\sin{\theta(t)\approx \theta(t)}$. Therefore, Equation (\ref{ph-equ-expan}) gives:

\begin{equation}
s_{pm}(t)=A_{pm} \cos{(\omega_c t)}-A_{pm}\sin{(\omega_c t)}a_m\sin{\left(\omega_m t\right)}.\label{fm-equ-expan2}
\end{equation}

Equation (\ref{fm-equ-expan2}) resembles Equation (\ref{two-waves}), they are similar in the sence that both are the product of two sinusoidal functions with different frequencies. So beginning with a phase modulation equation we can deduce, under certain approximations, an equation similar to Equation (\ref{two-waves}). This allows us to conclude as higly probable that the phase delay that is observed in Figure 1 is due to a phase modulation. This phase and amplitude modulation is also observed in Figure 2.

The preceding paragraps give a plausible explanation of the phase delay observed in the plots. This explanation is based on an implicit  similarity between Equations (\ref{fm-equ-expan2}) and (\ref{two-waves}). However, if there exist a phase modulation then there has to be a way to express Equation (\ref{ecu1}) in the form that Equation (\ref{ph-equation}) has. That is to say, we need to be able to write Equation (\ref{ecu1}) in the following two ways: 
\vspace{.5cm}
\begin{widetext}
\begin{equation}
y(t)=2A \cos\left(\frac{(\omega_2+\omega_1)t}{2}\right)\cos\left(\frac{(\omega_2-\omega_1)t}{2}\right)=A_{pm}\cos{\left[\omega_ct+ a_m\sin{(\omega_m t)}\right]}. \label{pm-way}
\end{equation}
\end{widetext}
\vspace{.5cm}
 Fortunatelly, this has been done long time ago. At least since the time where John William Strutt, 3rd Baron Rayleigh (a British scientist) lived. At the end of the 19 century Strutt wrote a book titled The Theory of Sound \cite{strutt}, in this book (see page 23) he wrote the sum of two waves similar to the last right part of Equation (\ref{pm-way}). The equation that he wrote is:
 
\begin{equation}
 u=r\cos\left(2\pi mt-\theta\right),\label{strutt1}
\end{equation}

where
\begin{equation}
r^2=a^2+a'^{2}+2aa'\cos\left(2\pi(m-n)t+\epsilon'-\epsilon \right),\label{strutt2}
\end{equation}
 and
\begin{equation}
\tan\theta=\frac{a\sin\epsilon+a'\sin\left(2\pi(m-n)t+\epsilon' \right)}{a\cos\epsilon+a'\cos\left(2(m-n)t+\epsilon \right)},\label{strutt3}
 \end{equation}
where $a$ and $a'$ are the amplitude of the two waves, $\epsilon$ and $\epsilon'$ are the initial phase. This equation can be calculated using the phasor method. So chosing $a=a'$ and $\epsilon=\epsilon'=0$ whe fulfill the initial condition of our problem.

Additionally, we have found a book that has an equation similar to Equation (\ref{strutt1}) \cite{rossing}. The authors of this book, Rossing and Fletcher \cite{rossing}, write the superposition of two waves as:
\begin{equation}
\widetilde{x}=A(t)e^{j(\omega_1t+\phi(t))},\label{ross}
\end{equation}
where $j$ is the imaginary number, and $A(t)$ and $\phi(t)$ are similar to Equations (\ref{strutt2}) and (\ref{strutt3}) respectively, see page $9$ in reference \cite{rossing}. Notably, these authors, clearly state that the resulting vibration has ``both amplitude and phase varying slowly''.

  In conclusion, the sum of two waves can be writing in the usual form that a phase modulation equation has, that is the far rigth of Equation (\ref{pm-way}) is true. Therefore, Equation (\ref{strutt1}) (and (\ref{ross})) allow us to conclude that effectively \textbf{there is a phase modulation} - at the same time that \textsl{there is an amplitude modulation} - when we sum two (or $N$) waves of nearly the same frequency.

Now, to finish this subsection, we will draw a historical issue about these points. It is interesting to notice that Equation (\ref{strutt1}) was no noticed before by many texbook's authors, including famous books like that of reference \cite{halliday}. On the other hand, it seems that people working in Sound uses this equation as a teaching tool. Then, the historical problem here is: Which book and author first published Equation (\ref{strutt1})?, Why scientist working in the subject matters of Waves and Optics have no noticed this equation before?.

\subsection{The diference between frequencies as a constant}
Instead of grouping the arithmetic sequence as we did in the the first paragrap (fifteen row) of Section II, we can group it by pairs as $100-50=99-49=98-48=\cdots52-2=51-1=50$. That is, the difference between numbers is a constant.  In this case, their sum descent by two in the interval between $[150,52]$ as follows: $100+50=150, 99+49=148, 98+48=146,\cdots 52+2=54, 51+1=52$.

Then, instead of Equation (\ref{groupingsum}) we can group together by pairs the wave's sum as follows:
\begin{eqnarray}
	y_T(t)=\bigg\{&&\left[\cos(\omega_{100}t)+\cos(\omega_{50}t)\right]+\left[\cos(\omega_{99}t)+\cos(\omega_{49}t)\right]+\nonumber\\ &&\cdots+	\left[\cos(\omega_{51}t)+\cos(\omega_{1}t)\right]\bigg\}.\label{groupdifferen}
\end{eqnarray}
Now, using the trigonometric identity given by equation (\ref{trigoidentity}) we have:
\begin{eqnarray}
	y_T(t)&&=\Bigg\{\left[
	\cos\left(\frac{(\omega_{100}+\omega_{50})t}{2}\right)\cos\left(\frac{(\omega_{100}-\omega_{50})t}{2}\right)\right]+\nonumber\\
	&&\left[\cos\left(\frac{(\omega_{99}+\omega_{49})t}{2}\right)\cos\left(\frac{(\omega_{99}-\omega_{49})t}{2}\right)\right]+\nonumber\\
	&&\cdots+
\left[\cos\left(\frac{(\omega_{51}+\omega_{1})t}{2}\right)\cos\left(\frac{(\omega_{51}-\omega_{1})t}{2}\right)\right]\Bigg\}\label{substrigo2}
\end{eqnarray}

Now, as $\omega_N=\omega_1+(N-1)\delta\omega$ then $\omega_{100}-\omega_{50}=\omega_{99}-\omega_{49}=\cdots=\omega_{51}-\omega_{1}=50\delta\omega$; also we have that $\omega_{100}+\omega_{50}=2\omega_1+148\delta\omega$, $\omega_{99}+\omega_{49}=2\omega_1+146\delta\omega$, $\cdots$, $\omega_{51}+\omega_{1}=2\omega_1+50\delta\omega$. Therefore,
\begin{eqnarray}
  \cos\left(\frac{(\omega_{100}-\omega_{50})t}{2}\right)=\cos\left(\frac{(\omega_{99}-\omega_{49})t}{2}\right)=\cdots\nonumber\\
  =\cos\left(\frac{(\omega_{51}-\omega_1)t}{2}\right)=\cos\left(\frac{50\delta\omega t}{2}\right),
\end{eqnarray}
so, in Equation (\ref{substrigo2}) we can factorize the cosines that have the same argument $(50\delta\omega t/2)$:
\begin{eqnarray}
y_T(t)=	\cos\left(25\delta\omega t\right)\Bigg\{\cos\left(\frac{(2\omega_1+148\delta\omega)t}{2}\right)+\nonumber\\
\cos\left(\frac{(2\omega_{1}+146\delta\omega)t}{2}\right)+\cdots+
\cos\left(\frac{(2\omega_{1}+50\delta\omega)t}{2}\right)\Bigg\}.\label{resta-factory}
\end{eqnarray}

In a further step, we can group by pairs cosines functions that are inside the braces in Equation (\ref{resta-factory}) in such a way that the difference between the arguments of the cosine function are equal. This will provide twenty five cosine functions, and the procedure can be followed to the step where you find the shortes expression. However, we will not carried ou this procedure, instead in the next subsection we focouses in the case where we have only ten waves.

\subsubsection{Example 2}

In this subsubsection we are going to study the case of the superposition of ten waves. In this case the arithmetic sequence is $\omega_1, \omega_{2}=\omega_1+\delta\omega, \omega_{3}=\omega_1+2\delta\omega, \omega_{4}=\omega_1+3\delta\omega, \omega_{5}=\omega_1+4\delta\omega, 
\omega_{6}=\omega_1+5\delta\omega, \omega_{7}=\omega_1+6\delta\omega, \omega_{8}=\omega_1+7\delta\omega, \omega_{9}=\omega_1+8\delta\omega,
 \omega_{10}=\omega_1+9\delta\omega$. The total superposition will be -where we have grouped by pairs the cosines functions in a convenient way-:
 
\begin{eqnarray}	
&&y_T(t)=A\bigg\{\left[\cos(\omega_{10}t)+\cos(\omega_5t)\right]+\nonumber\\
&&\left[\cos(\omega_{9}t)+\cos(\omega_{4}t)\right]+\left[\cos(\omega_{8}t)+\cos(\omega_{3}t)\right]\nonumber\\
&&+\left[\cos(\omega_{7}t)+\cos(\omega_{2}t)\right]	+\left[\cos(\omega_{6}t)+\cos(\omega_{1}t)\right]\bigg\}.\label{}
\end{eqnarray}

Following the same steps of the Example 1 in Section II, we arrive to the following final result:  

\begin{eqnarray}
y_T(t)=2A\cos\left(\frac{5\delta\omega}{2}t\right)&&\Bigg\{4\cos^2(2\delta\omega t)\cos\left[(4w_1+20\delta\omega) t\right]+\nonumber\\
&&\cos\left[\frac{2\omega_1+5\delta\omega}{2}t \right] \Bigg\}
\end{eqnarray}

\section{Conclusions}
 In this work we present the well known Gauss method to sum an arithmetic secuence, but now as an useful aid to add $N$ waves. This method is simple, easy to apply and avoids students to use complex number and remember complex equations reported in the literature of waves. This method can be used to teach this subject matter since the beginning of the syllabus.
 
 Most important, results observed in the plots shows a phase delay in the superposition of waves not previously reported. We have shown that togheter with an amplitude modulation there must exist a \textbf{phase modulation}. 
 
 The amplitude modulation produces the phenomenom of beating. We think that the \textbf{phase modulation} produces a phenomenon that can not be noticed by human detectors (the ears), but probably this phenomenon could be recorded by other detector system.

\begin{acknowledgments}
L. M. Ar\'evalo Aguilar thanks the support of Consejo Nacional de Ciencia y T\'ecnologia (CONACYT) under grant No. 117319. 
C. Robledo-S\'anchez, M. L. Arroyo Carrasco and M. Otero thanks the support from CONACYT.\end{acknowledgments}


\end{document}